\documentclass[aps,pra,twocolumn,amsmath,amssymb,showpacs]{revtex4-1}

\usepackage{graphicx}
\usepackage{dcolumn}
\usepackage{bm}
\usepackage{CJK}
\usepackage{color}

\def\hatd#1{\hat{#1}^\dagger}

\def\ket#1{\left|{#1}\right\rangle}
\def\braket#1#2{\left\langle{{#1}}\mathrel{\left|{\vphantom{{#1}{#2}}}\right.\kern-\nulldelimiterspace}{{#2}}\right\rangle}

\def\Jp{J_\parallel}
\def\Jv{J_\perp}
\def\romanNum#1{\romannumeral#1}
\def\RomanNum#1{\textrm{\uppercase\expandafter{\romannumeral#1}}}

\begin{document}


\title{Cluster Gutzwiller study of Bose-Hubbard ladder: ground-state phase diagram and many-body Landau-Zener dynamics}

\author{Haiming Deng$^{1,2,3}$}
\author{Hui Dai$^{1}$}
\author{Jiahao Huang$^{1,2}$}
\author{Xizhou Qin$^{1,2}$}
\author{Jun Xu$^{1,2,4}$}
\author{Honghua Zhong$^{1,2,5}$}
\author{Chunshan He$^{1}$}

\author{Chaohong Lee$^{1,2}$}
\altaffiliation{Corresponding author. Email: chleecn@gmail.com}

\affiliation{$^{1}$State Key Laboratory of Optoelectronic Materials and Technologies, School of Physics and Engineering, Sun Yat-Sen University, Guangzhou 510275, China}
\affiliation{$^{2}$Institute of Astronomy and Space Science, Sun Yat-Sen University, Guangzhou 510275, China}
\affiliation{$^{3}$College of Electronic Information and Electrical Engineering, Xiangnan University, Chenzhou 423000, China}
\affiliation{$^{4}$Center of Experimental Teaching for Common Basic Courses, South China Agriculture University, Guangzhou 510642,China}
\affiliation{$^{5}$Department of Physics, Jishou University, Jishou 416000, China}

\date{\today}

\begin{abstract}
  We present a cluster Gutzwiller mean-field study for ground states and time-evolution dynamics in the Bose-Hubbard ladder (BHL), which can be realized by loading Bose atoms in double-well optical lattices.
  In our cluster mean-field approach, we treat each double-well unit of two lattice sites as a coherent whole for composing the cluster Gutzwiller ansatz, which may remain some residual correlations in each two-site unit.
  For a unbiased BHL, in addition to conventional superfluid phase and integer Mott insulator phases, we find that there are exotic fractional insulator phases if the inter-chain tunneling is much stronger than the intra-chain one.
  The fractional insulator phases can not be found by using a conventional mean-field treatment based upon the single-site Gutzwiller ansatz.
  For a biased BHL, we find there appear single-atom tunneling and interaction blockade if the system is dominated by the interplay between the on-site interaction and the inter-chain bias.
  In the many-body Landau-Zener process, in which the inter-chain bias is linearly swept from negative to positive or vice versa, our numerical results are qualitatively consistent with the experimental observation [Nat. Phys. \textbf{7}, 61 (2011)].
  Our cluster bosonic Gutzwiller treatment is of promising perspectives in exploring exotic quantum phases and time-evolution dynamics of bosonic particles in superlattices.
\end{abstract}


\maketitle

\section{Introduction}\label{Sec1}

The unprecedented experimental techniques of manipulating and detecting ultracold atoms in optical lattices~\cite{Greiner2002, Bloch2005} provide an ideal testing ground to investigate Bose-Hubbard (BH) models~\cite{Greiner2002, Bloch2005, morsch2006, Lewenstein2007, Bloch2008, Will2010, Polkovnikov2011, David2011}.
The remarkable cleanness and high tunability of ultracold atomic systems allow one to explore various many-body quantum phenomena in BH models~\cite{Paredes2004,Kinoshita2004,Hofferberth2007}.
For an example, the experimental realization of the one-dimensional (1D) atomic Hubbard model~\cite{Stoferle2004} provides new opportunities to exploring quantum statistical effects and strong correlation effects in low-dimensional quantum many-body systems~\cite{guan2013}.
Quantum dynamics as well as quantum phase transition between superfluid (SF) phase and Mott insulator (MI) phase in BH models are of great interests and have been widely investigated~\cite{Jaksch1998,Freericks1994,Grusdt2013,Muth2008,BLChen2010,Danshita2007,Kollath2007,Lee2004}.

In recent, by loading ultracold Bose atoms into a double-well optical lattice potential, the Bose-Hubbard ladder (BHL) had been realized and the many-body Landau-Zener (LZ) dynamics has been explored~\cite{Chen2011}.
Different from the single-particle LZ process, the breakdown of adiabaticity in the inverse sweeping from the highest excited state had been observed in the many-body LZ process of the BH ladder.
This experiment has stimulated extensive investigation of both stationary and dynamic behaviors in the BHL via different theoretical methods, such as, full diagonalization method~\cite{Tschischik2012} and time-dependent density-matrix renormalization group (\emph{t}-DMRG) technique~\cite{Kasztelan2011,pandey,Keles2015}.
However, the full diagonalization and \emph{t}-DMRG methods should cost a huge number of computational resources.

To simulate the BHL with less computational resources, the bosonic Gutzwiller method~\cite{Rokhsar1991,Krauth1992} is an alternative option.
Although the Gutzwiller method has common restrictions of the mean-field methods, it has provided versatile applications in qualitative calculations of both stationary states and time-evolution dynamics.
In recent, cluster bosonic Gutzwiller methods~\cite{Buonsante2004, Pisarski2011, McIntosh2012, Yamamoto2012a, Yamamoto2012b, Luhmann2013} have been developed by coupling multi-site clusters rather than single sites with the mean field.
By employing the cluster Gutzwiller method, some properties of many-body LZ phenomena in repulsive BHL~\cite{Benitez2012} and  some quantum phases in attractive BHL~\cite{singh2014} have been explored.
In Ref.~\cite{Benitez2012}, the phase diagram and LZ dynamics for fixed average number of particles per site have been shown. However, it does not discuss how the phase diagram and LZ dynamics depend on the chemical potential.
In the large-size multi-site Gutzwiller method~\cite{singh2014}, the three-body constraint has been imposed to each lattice site. In a realistic experimental system, the number of particles in each lattice site may break this constraint.
In addition, the ground-state phase diagrams of BH systems have been obtained by the analytical mean-field approach~\cite{Buonsante2005LP}, the cell strong-coupling perturbation technique~\cite{Buonsante2005PRA} and the composite boson mean-field theory~\cite{Huerga2013PRL, Zhao2014JPCM} etc.

In this article, we present a cluster Gutzwiller mean-field study for the ground-state phase diagram and many-body LZ dynamics of a BHL.
In our mean-field treatment, we regard each double-well unit of two lattice sites as a coherent whole for composing the cluster Gutzwiller ansatz, which remains some residual inter-site correlations in each double-well unit.
For a unbiased BHL, in addition to superfluid (SF) and integer Mott insulator (MI) phases which may be found by single-site Gutzwiller treatment, we find that there exist exotic fractional insulator phases if the inter-chain tunneling is much stronger than the intra-chain one.
The exotic fractional insulator phase in BHL is similar to the rung-Mott phase in hard-core BH system~\cite{singh2014,carrasquilla2011}.
We also obtain the phase diagram for the asymmetry BHL, which have not yet been reported.
In further, by linearly sweeping the inter-chain bias from negative to positive or vice versa, we analyze many-body LZ dynamics in the system and confirm the existence of adiabaticity breakdown.

This article is organized as follows. In Sec.~\ref{Sec2}, we give the physical model and discuss its realization. In Sec.~\ref{Sec3}, we present the cluster Gutzwiller mean-field method and obtain the ground-state phase diagram for both symmetric and asymmetric BHLs. In Sec.~\ref{Sec4}, we study the many-body LZ dynamics and show the adiabaticity breakdown. At last, in Sec.~\ref{Sec5}, we briefly summarize and discuss our results.

\section{Model}\label{Sec2}

\begin{figure}[!htb]
  \includegraphics[width=\columnwidth]{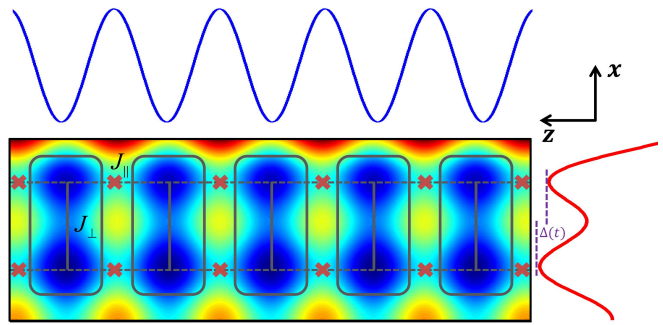}
  \caption{\label{Fig_schematic} Schematic diagram of the superlattice potential~\eqref{Superlattice} projected onto the x-z plane.
  The time-dependent bias $\Delta(t)$ can be achieved by ramping up the lattices to obtain the tilted potential along the x-axis.
  Here, the parameters are set as $\lambda_{z}=\lambda_{xl}=2\lambda_{xs}$ and $V_z=1.5V_{xl}=1.5V_{xs}$.
  The potential along the x-axis within a period of $\lambda_{xl}$ is an asymmetric double-well potential and the potential along the z-axis is a standing-wave potential.
  The two lattice sites in each double-well potential are packed as a cluster, which are depicted by the gray rectangles.
  The clusters are decoupled by applying the Gutzwiller mean-field treatment, in which the crosses stand for the decoupling between neighboring clusters and the gray dashed and solid lines respectively denote the intra- and inter-chain tunneling.}
\end{figure}

We consider an ensemble of Bose atoms confined within a double-well superlattice potential,
\begin{eqnarray}\label{Superlattice}
  V(x,z) &=& V_{xl}\sin^2(2\pi x/\lambda_{xl})+ V_{xs}\sin^2(2\pi x/\lambda_{xs}) \nonumber\\
  &+& V_{z}\sin^2(2\pi z/\lambda_{z}),
\end{eqnarray}
where the first and second terms are generated by superimposing two standing-wave lasers along the x-direction with wavelengths $\lambda_{xl}$ and $\lambda_{xs}$.
The two potential depths $V_{xs}$ and $V_{xl}$ are determined by the laser intensities.
To form the double-well lattice potential, the wavelengths are set to be $\lambda_{xl}=2\lambda_{xs}$.
The last term describes a lattice potential along the z-direction with the wavelength $\lambda_{z}$ and the depth $V_{z}$.
During the experiment, the energy difference between the lattices in each double-well unit can be ramped up or down with time.
The schematic diagram for the double-well lattices is shown in Fig.~\ref{Fig_schematic}.

If the barriers between neighboring double-well units along the x-direction is sufficiently high, the system can be described by several parallel BHLs with ignorable inter-ladder couplings.
The Hamiltonian for a single BHL reads as,
\begin{eqnarray}\label{Eq_Hamiltonian}
  \hat H(t)&=&-\Jp\sum_{\left\langle{jk}\right\rangle\sigma}\hatd{b}_{j\sigma}\hat{b}_{k\sigma}
  -\Jv\sum_j\left(\hatd{b}_{jL}\hat{b}_{jR}+\mathrm{h.c.}\right)\nonumber\\
  &&+\frac{U}{2}\sum_{j\sigma}\hat{n}_{j\sigma}\left(\hat{n}_{j\sigma}-\hat{1}\right)
  -\frac{\Delta(t)}{2}\sum_j(\hat{n}_{jR}-\hat{n}_{jL})\nonumber\\
  &&-\mu\sum_{j\sigma}\hat{n}_{j\sigma},
\end{eqnarray}
where, $\left\langle{jk}\right\rangle$ indicates the summation comprising all nearest neighboring sites in the same chain and the index $\sigma=(L,R)$ denotes the left or right chain.
The symbols $\hatd{b}_{j\sigma}$ ($\hat{b}_{j\sigma}$) creates (annihilates) a Bose atom on the $j$-th lattice of the $\sigma$-chain, and $\hat{n}_{j\sigma}=\hatd{b}_{j\sigma}\hat{b}_{j\sigma}$ stands for the atomic number.
The parameters $\Jp$ and $\Jv$ are the intra- and inter-chain nearest-neighbor hopping strengths, respectively.
The on-site interaction $U$ is determined by the s-wave scattering lengthes and the chemical potential $\mu$ determines the particle filling.

To characterize different regimes of the BHL, we introduce the ratio between the intra- and inter-chain hopping strengths $\beta=\Jp/\Jv$.
If $\beta \ll 1$, the intra-chain tunneling is ignorable and the ladder system can be regarded as isolated double-wells.
However, if $\beta \gg 1$, the intra-chain tunneling becomes dominant, the system can be treated as two decoupled single BH chains.
The parameter $\Delta(t)$ represents the inter-chain energy bias.

\section{Ground-state phase diagram}\label{Sec3}

In this section, we show how to obtain the ground-state phase diagram via the cluster Gutzwiller mean-field treatment. In the first subsection, we describe the cluster Gutzwiller mean-field approach for the BHL. Then in the second subsection, we present the self-consistent procedure for determining ground states. In the last subsection, we give the ground-state phase diagram.

\subsection{Cluster Gutzwiller mean-field approach}

The standard Gutzwiller method assumes the wavefunction of the whole system as a product state of single-site wavefunctions. By implementing the standard Gutzwiller procedure, the BH model is decoupled as single sites which couple with surround sites via their average mean fields. In further, attribute to the equivalence of all lattice sites in the model, one can replace the mean fields of surround sites with the mean field of the site itself and so that the mean-field version for the original Hamiltonian can be written as a sum of single-site terms.

In the following, the cluster Gutzwiller mean-field approach is an extension of the single-site Gutzwiller mean-field approach. As shown in Fig.~\ref{Fig_schematic}, the decoupling holds for each double-well cluster which includes one lattice site in the left chain and one lattice site in right chain. Therefore all clusters are equivalent and the state for the whole system is written as a product state of the single-cluster states which remains the correlations between lattice sites in the same cluster.
Unlike the single-site Gutzwiller approach, in which all tunneling terms are decoupled, the cluster Gutzwiller approach keeps the intra-cluster tunneling terms and only decouple the inter-cluster tunneling terms.
By using the mean-field treatment, the inter-cluster tunneling terms are decoupled as
\begin{eqnarray}\label{Eq_meanfield}
  \hatd{b}_{j\sigma}\hat{b}_{k\sigma}&\approx&
  \hatd{b}_{j\sigma}\langle\hat{b}_{k\sigma}\rangle +\langle\hatd{b}_{j\sigma}\rangle\hat{b}_{k\sigma}
  -\langle\hatd{b}_{j\sigma}\rangle\langle\hat{b}_{k\sigma}\rangle
  \nonumber\\
  &=&\hatd{b}_{j\sigma}\varphi_{k\sigma} +\varphi_{j\sigma}^*\hat{b}_{k\sigma} -\varphi_{j\sigma}^*\varphi_{k\sigma},
\end{eqnarray}
where $\varphi_{k\sigma}=\langle \hat b_{k\sigma} \rangle$ and the high-order fluctuations $\delta_{\hatd{b}_{j\sigma}} \delta_{\hat{b}_{k\sigma}}
=(\hatd{b}_{j\sigma}-\langle\hatd{b}_{j\sigma}\rangle)(\hat{b}_{k\sigma} -\langle\hat{b}_{k\sigma}\rangle)$ are neglected. Therefore the original Hamiltonian~(\ref{Eq_Hamiltonian}) is decoupled as
\begin{equation}\label{Eq_MFHam}
  \hat{H}^\mathrm{MF}=\sum_{j}\hat{H}_j^\mathrm{MF},
\end{equation}
with the single-cluster mean-field Hamiltonian
\begin{eqnarray}\label{Eq_MFHam_j0}
  \hat{H}_j^\mathrm{MF}&=&-\Jp\sum_{\sigma,k=j\pm1}\left(
  \varphi_{k\sigma}\hatd{b}_{j\sigma}+\varphi_{k\sigma}^*\hat{b}_{j\sigma}
  -\mathrm{Re}\left[\varphi_{j\sigma}^*\varphi_{k\sigma}\right]\right)\nonumber\\
  &&-\Jv\left(\hatd{b}_{jL}\hat{b}_{jR}+\mathrm{h.c.}\right)
  +\frac{U}{2}\sum_{\sigma}\hat{n}_{j\sigma}\left(\hat{n}_{j\sigma}-\hat{1}\right)\nonumber\\
  &&-\frac{\Delta}{2}(\hat{n}_{jR}-\hat{n}_{jL})-\mu\sum_{\sigma}\hat{n}_{j\sigma}.
\end{eqnarray}

Making use of the Gutzwiller ansatz, the state for the whole system can be expressed as a product state of single-cluster states,
\begin{equation}\label{Eq_GA}
  \ket{\Psi^\mathrm{GA}}=\prod_{j}\ket{\Psi}_j,
\end{equation}
where the state for the $j$-th cluster $|\Psi\rangle_{j}$ can be expanded as
\begin{equation}\label{Eq_GA_j}
  \ket{\Psi}_j=\sum_{N=0}^{N_\mathrm{max}}\sum_{m=-N}^{N}f_{N,m}^{(j)}\ket{N,m}_j.
\end{equation}
with $\ket{N,m}_j$ denoting the state basis for the $j$-th cluster.
Here, $N=N_L+N_R$, $m=N_L-N_R$, $N_L$ ($N_R$) stands for the number of particles in the left (right) chain, the probability amplitudes $f_{N,m}^{(j)}$ are complex numbers, and $N_\mathrm{max}$ is the truncation of the maximum particle number.

Obviously, it is easy to find that the eigenequation $\hat H^\mathrm{MF}\ket{\Psi^\mathrm{GA}}=E\ket{\Psi^\mathrm{GA}}$ for the whole system is equivalent to the single-cluster eigenequation
\begin{equation}\label{Eq_eigen_MF}
  \hat{H}_j^\mathrm{MF}\ket{\Psi}_j=E_j\ket{\Psi}_j
\end{equation}
with $E=\sum_j{E_j}$.
By substituting the single-cluster state~\eqref{Eq_GA_j} into the single-cluster eigenequation~\eqref{Eq_eigen_MF}, we obtain
\begin{eqnarray}\label{Eq_meanfield_Ham}
  E_{j}f_{N,m}^{(j)}=
  &-&\frac{\Jp}{\sqrt{2}}\phi_{jL}\sqrt{N+m}f_{N-1,m-1}^{(j)}\nonumber\\
  &-&\frac{\Jp}{\sqrt{2}}\phi_{jR}\sqrt{N-m}f_{N-1,m+1}^{(j)}\nonumber\\
  &-&\frac{\Jp}{\sqrt{2}}\phi_{jL}^*\sqrt{N+m+2}f_{N+1,m+1}^{(j)}\nonumber\\
  &-&\frac{\Jp}{\sqrt{2}}\phi_{jR}^*\sqrt{N-m+2}f_{N+1,m-1}^{(j)}\nonumber\\
  &+&\Jp\mathrm{Re}\left[\varphi_{jL}^*\phi_{jL}+\varphi_{jR}^*\phi_{jR}\right]f_{N,m}^{(j)}\nonumber\\
  &-&\frac{\Jv}{2}\sqrt{N+m}\sqrt{N-m+2}f_{N,m-2}^{(j)}\nonumber\\
  &-&\frac{\Jv}{2}\sqrt{N+m+2}\sqrt{N-m}f_{N,m+2}^{(j)}\nonumber\\
  &+&\left[\frac{U}{4}\left(N^2+m^2-2N\right) +\frac{\Delta}{2}{m}-\mu{N}\right]f_{N,m}^{(j)}.\nonumber\\
\end{eqnarray}
Here, the order parameters are quantum expectation values of bosonic annihilation operators, i.e. $\varphi_{jL}=\langle \Psi^{GA}|\hat b_{jL}|\Psi^{GA} \rangle$ and $\varphi_{jR}=\langle \Psi^{GA}|\hat b_{jR}|\Psi^{GA} \rangle$.
After some mathematical calculation, we have
\begin{eqnarray}
  \varphi_{jL}&=&\sum_{N,m}\sqrt{\frac{N+m+2}{2}}f_{N,m}^{(j)*}f_{N+1,m+1}^{(j)}\label{Eq_ord para1}, \\
  \varphi_{jR}&=&\sum_{N,m}\sqrt{\frac{N-m+2}{2}}f_{N,m}^{(j)*}f_{N+1,m-1}^{(j)}\label{Eq_ord para2}.
\end{eqnarray}
For convenience, we define $\phi_{jL}=\varphi_{(j+1)L}+\varphi_{(j-1)L}$ and $\phi_{jR}=\varphi_{(j+1)R}+\varphi_{(j-1)R}$.

\subsection{Self-consistent procedure for determining ground states}

As the single-cluster Hamiltonian~\eqref{Eq_MFHam_j0} depends on the mean fields, one has to implement self-consistent procedure for determining the mean fields and the ground states.
Given the parameters $U$, $\Jp$, $\Jv$, $\mu$ and $\Delta$, one can obtain the ground state from the single-cluster eigenequation~\eqref{Eq_eigen_MF} and the self-consistent relations $\varphi_{\sigma}=\langle \hat b_{\sigma} \rangle$.
In Fig.~2, we show the key steps of the self-consistent procedure for determining ground states.

(\romanNum{1}) Initialize $\varphi_{\sigma}=0$, $\varphi_{\sigma}'=\varphi_{\sigma}$ and set the trial ground state energy $E_{GS}^{min}$ an arbitrary value.

(\romanNum{2}) Substitute $\varphi_{\sigma}$ into the single-cluster Hamiltonian, and diagonalize the Hamiltonian to obtain its ground state with eigen-energy $E_{GS}$.

(\romanNum{3}) If $E_{GS}<E_{GS}^{min}$, replace $E_{GS}^{min}$ and $\varphi_{\sigma}'$ with $E_{GS}$ and $\varphi_{\sigma}$, respectively. Otherwise, let $\varphi_{\sigma}=\varphi_{\sigma}+\Delta\varphi_{\sigma}$ and implement step~(\romanNum{2}) again.

(\romanNum{4}) Repeat Steps~(\romanNum{2}) and~(\romanNum{3}) until $\varphi_{\sigma}\ge \sqrt{N_{max}}$.

(\romanNum{5}) Set $\varphi_\sigma^\RomanNum{1}=\varphi_{\sigma}'$ and calculate the ground state $\ket{\mathrm{GS}_\RomanNum{1}}$ from the single-cluster eigenequation.

(\romanNum{6}) Calculate the order parameters $\varphi_\sigma^\RomanNum{2}$ for the ground state $\ket{\mathrm{GS}_\RomanNum{1}}$.

(\romanNum{7}) Compare $\varphi_\sigma^\RomanNum{1}$ and $\varphi_\sigma^\RomanNum{2}$, if $\left|\varphi_\sigma^\RomanNum{1}-\varphi_\sigma^\RomanNum{2}\right|<\epsilon$ (where $\epsilon$ is pre-given tolerance), then output $\ket{\mathrm{GS}_\RomanNum{1}}$ as the ground state. Otherwise, set $\varphi_\sigma^\RomanNum{1}=\varphi_\sigma^\RomanNum{2}$ and return to step~(\romanNum{5}).

Through the procedure from step (\romanNum{1}) to step (\romanNum{4}), one can numerically minimize the system energy with respect to the order parameters in the interval of $\varphi_{\sigma}\in[0,\sqrt{N_\mathrm{max}}]$.
Usually, the steps~(\romanNum{5}-\romanNum{7}) are the so-called self-consistent procedure.

\begin{figure}[!htp]
  \includegraphics[width=\columnwidth]{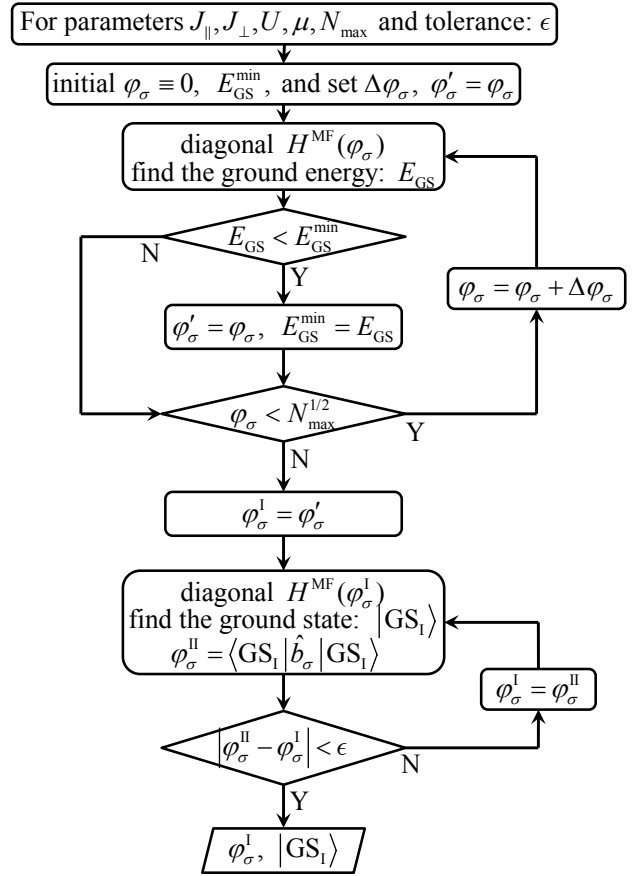}
  \caption{\label{Fig_GS_procedure} The numerical self-consistent procedure for solving the single-cluster eigenequation.}
\end{figure}

\subsection{Phase diagram}

In this subsection, we show the ground-state phase diagram for the BHL. Our cluster mean-field approach can be applied to both symmetric and asymmetric systems. Below, we first consider the symmetric system with no inter-chain bias (i.e. $\Delta(t)=0$), then consider the asymmetric cases with nonzero inter-chain bias $\Delta$.

\begin{figure*}[!htp]
  \includegraphics[width=17.2 cm]{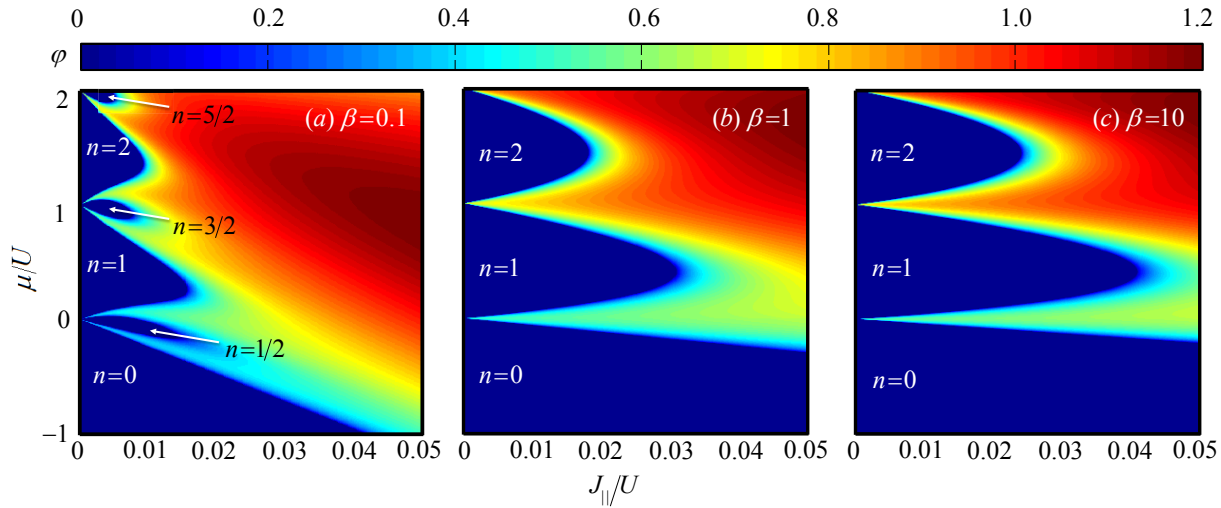}
  \caption{\label{Fig_phase_diagrams} (Color online) The ground-state phase diagram for the symmetric Bose-Hubbard ladder ($\Delta=0$) with the on-site interaction $U=1$ and different values of $\beta=\Jp/\Jv$: (a) $\beta=0.1$, (b) $\beta=1$, and (c) $\beta=10$.
  In our calculation, we set the truncation of maximum particle number $N_\mathrm{max}=6$ whose validity has been numerically verified.
  The blue areas are the insulator phases with zero order parameter $\varphi_j=0$ with $n$ denoting the filling number (the average atomic number per lattice site).
  The Mott insulator (MI) lobes have integer filling numbers, while the loophole insulator (LI) phases have half-integer filling numbers.
  The regions outside the blue areas are the superfluid (SF) phases with nonzero order parameters $\varphi_j\neq 0$.
  }
\end{figure*}

In Fig.~\ref{Fig_phase_diagrams}, we show the ground-state phase diagram for symmetric BHL with $\Delta(t)=0$ and different ratios $\beta=\Jp/\Jv$.
Due to the absence of asymmetry, the two chains are completely equivalent and the order parameters of both chains are always equal $\varphi_{jL}=\varphi_{jR}=\varphi_{j}$.
Therefore, it is enough to give the phase diagram via analyzing the order parameter for one of the two chains.
The ground states sensitively depend on the chemical potential $\mu$, the on-site interaction $U$, the intra-chain hopping $\Jp$ and the inter-chain hopping $\Jv$.

Usually, determined by the superfluid order parameter, the BH systems have two typical phases: (i) the superfluid (SF) phase of nonzero order parameter and (ii) the Mott insulator (MI) phase of zero order parameter and integer filling number.
For our atomic BHL system, when the inter-cluster hopping $\Jp$ is sufficiently strong, the atoms can move freely between neighboring double-well clusters and there appears a SF along the chain direction.
In contrast, when the on-site interaction $U$ becomes sufficiently strong, the atoms are localized in each cluster and there is no SF along the chain direction.
The chemical potential $\mu$ controls the filling number, i.e. the average atomic number per site.

Under the condition of $\beta=\Jp/\Jv \gg 1$, i.e. the intra-chain tunneling is much stronger than the inter-chain tunneling, the BHL can be regarded as two decoupled chains and the corresponding phase diagram is almost as same as the one for a single BH chain. In Fig.~\ref{Fig_phase_diagrams} (c), we show the phase diagram for $\beta=10$. At the side of strong intra-chain tunneling, $\Jp /U \rightarrow +\infty$, the ground states are SF phases of nonzero order parameter. At the side of strong interaction, $\Jp /U \rightarrow 0$, there appear several integer MI lobes which has integer filling numbers per lattice site and zero order parameter. The blue region in the bottom corresponds to the vacuum state with no any atoms. The biggest lobe corresponds to the MI phase with definitely one atom ($n=1$) in each site and the smaller one stands for the MI phase of $n=2$. This phase diagram reminds us the one for the one-dimensional BH model~\cite{Fisher1989}.

The areas of MI lobes shrink if the ratio $\beta$ decreases, see Fig.~\ref{Fig_phase_diagrams}~(a-c). Qualitatively, the shrinking of MI lobes can be understood by the intra-chain tunneling assisted by the inter-chain tunneling.
When the ratio $\beta$ becomes very small, the inter-chain hopping $\Jv$ are much stronger than the intra-chain hopping $\Jp$, the areas of MI lobes shrink dramatically, and, interestingly, several loophole insulator (LI) phases of zero order parameters appear between the conventional MI lobes, see Fig.~\ref{Fig_phase_diagrams}~(a).

To distinguish the LI and MI phases, we calculate the filling numbers (the average atomic numbers per site) and find that the LI phases have half-integer filling numbers and while the MI phases have integer filling numbers.
The half-integer filling numbers mean that the total atomic numbers per cluster are odd integer numbers and the residual atom in each double-well cluster can freely move between the two wells of each cluster. In further, we calculate the intra-cluster first-order correlation $\textrm{Cor}_{\perp}^{(1)}=\left|\langle\hatd{b}_{jL}\hat{b}_{jR}\rangle\right|$ and find that the LI phases have nonzero $\textrm{Cor}_{\perp}^{(1)}$.

The appearance of the LI phases is a direct result of $U \gg\Jp$ and $\Jv \gg \Jp$.
As $U \gg \Jp$, the tunneling along the chain direction is suppressed and the order parameter vanishes. However, the atoms in each double-well cluster may still freely move between the two wells and so that the total atomic numbers per cluster may be odd integer numbers.
The insulator phases of fractional filling numbers have also been found in one-dimensional superlattice BH models \cite{Buonsante2004,Grusdt2013} via mean-field method, quantum Monte Carlo simulation and numerical density matrix renormalization group simulation.
Different from the one-dimensional superlattice BH chains \cite{Buonsante2004,Grusdt2013}, our ladder system includes two coupled one-dimensional BH chains and the coupling between different clusters are more complex.

\begin{figure*}[!htp]
  \includegraphics[width=17.2 cm]{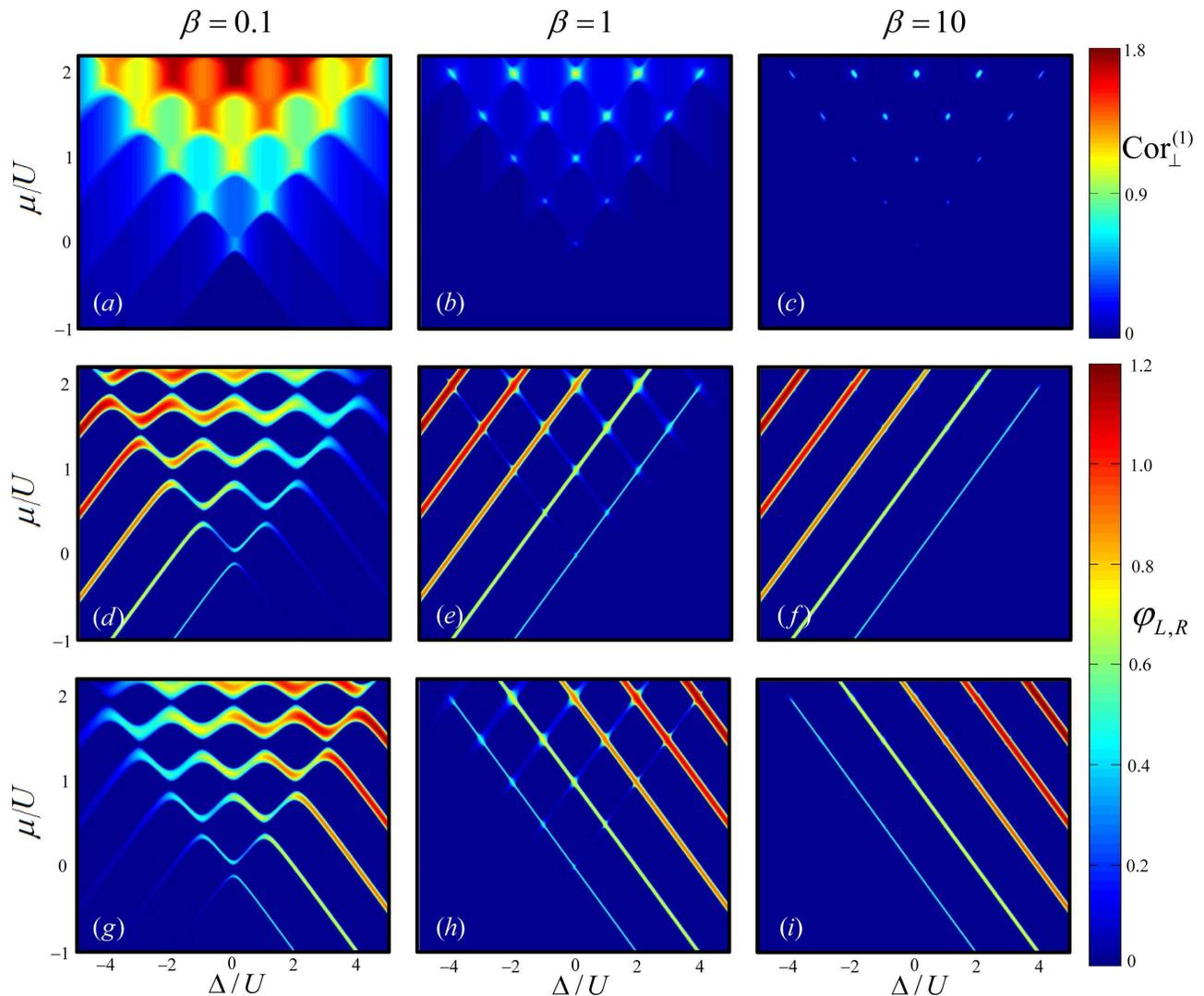}
  \caption{\label{Fig_phase_diagrams2} (Color online) The ground-state phase diagrams for the biased Bose-Hubbard ladder with different values of $\beta=\Jp/\Jv$ versus the bias $\Delta$. In our calculation, we fix the on-site interaction $U=1$, the intra-chain hopping $\Jp=0.01$, and set the truncation of maximum particle number $N_\mathrm{max}=6$. The first row shows the intra-cluster first-order correlation $\textrm{Cor}_{\perp}^{(1)}$, and the second and third rows respectively correspond to the order parameters $\varphi_{L,R}$ for the left and right chains. The hopping ratios are chosen as $\beta=0.1$ (the first column), $\beta=1$ (the second column) and $\beta=10$ (the third column), respectively.
  }
\end{figure*}

Now, we discuss the ground-state phase diagrams for BHL with nonzero bias $\Delta$. Due to nonzero bias $\Delta$, the two chains are no longer equivalent and so that the order parameters $\varphi_{jL}$ and $\varphi_{jR}$ for the left and right chains may have different values. In Fig.~\ref{Fig_phase_diagrams2}, we show the two order parameters ($\varphi_{jL}$, $\varphi_{jR}$) (the second and third rows) and the intra-cluster first-order correlation $\textrm{Cor}_{\perp}^{(1)}$ (the first row).

In our numerical simulation, by employing the cluster mean-field method presented in Subsection A, we consistently obtain the ground-state phase diagram in the $(\Delta/U, \mu/U)$-plane for the intra-chain hopping strength $\Jp=0.01$, the on-site interaction $U=1$, and different values of the hopping ratio $\beta=\Jp/\Jv=(0.1, 1, 10)$.
Our results show that, for given $\Jp$, $\beta$ and $\mu$, the order parameter for the left chain $\varphi_{jL}$ with bias $\Delta$ equals to the order parameter for the right chain $\varphi_{jR}$ with bias $-\Delta$.
Therefore, the ground-state phase diagrams of $\varphi_{jL}$ and $\varphi_{jR}$ are symmetric with each other about the axis $\Delta=0$ for given $\Jp$ and $\beta$.
Moreover, the intra-cluster correlation $\textrm{Cor}_{\perp}^{(1)}$ is also symmetric about the axis $\Delta=0$.

If the inter-chain tunneling is very weak, that is the hopping ratio $\beta \gg 1$, the BHL can almost be treated as two independent chains.
In the third column of Fig.~4, we show the inter-chain coherence $\textrm{Cor}_{\perp}^{(1)}$ and the order parameters ($\varphi_{jL}$, $\varphi_{jR}$) for $\beta=10$.
In phase diagrams of ($\varphi_{jL}$, $\varphi_{jR}$), there appear several parallel and equal-spaced SF stripes with nonzero order parameters $\varphi_{jL}$ or $\varphi_{jR}$, see Fig.~4 (f) and (i).
The SF stripes of $\varphi_{jL}$ become narrower and its corresponding values become smaller as the bias $\Delta$ increases from the negative to the positive side, see Fig.~4 (f).
While for the order parameter $\varphi_{jR}$, its SF stripes change oppositely as they are symmetric with the ones of the order parameter $\varphi_{jL}$ about the axis $\Delta=0$, see Fig.~4 (i).
The blue regions correspond to MI phases with zero order parameters $\varphi_{jL}$ or $\varphi_{jR}$.
For the inter-chain coherence $\textrm{Cor}_{\perp}^{(1)}$, nonzero values only appear in the vicinity surrounding some specific points, which form an inverted triangle lattice structure, see Fig.~4 (c).

The interplay between the inter-chain bias and the on-site interaction under weak hopping leads to the exotic single-atom tunneling [the bright spots in Fig.~4 (c)] and the interaction blockade [the blue regions in Fig.~4 (c)].
Under strong on-site interactions, that is $U\gg\Jp$ and $U\gg\Jv$, the system behavior can be understood by a perturbation picture.
As $U\gg\Jp$ and $U\gg\Jv$, the hopping terms can be regarded as perturbations and the dominant part of the BHL Hamiltonian reads as
\begin{eqnarray}\label{Eq_Hamiltonian_blockade}
  \hat H_d&=&\frac{U}{2}\sum_{j\sigma}\hat{n}_{j\sigma}\left(\hat{n}_{j\sigma}-\hat{1}\right)
  -\frac{\Delta}{2}\sum_j(\hat{n}_{jR}-\hat{n}_{jL})\nonumber\\
  &&-\mu\sum_{j\sigma}\hat{n}_{j\sigma}.
\end{eqnarray}
From the Fock states for a cluster, a single Fock state $|n_L,n_R\rangle$ corresponds to a MI phase of zero order parameters, the quasi-degeneracy of $|n_L,n_R\rangle$ and $|n_L+1,n_R\rangle$ will result a nonzero order parameter $\varphi_{jL}$, and the quasi-degeneracy of $|n_L,n_R\rangle$ and $|n_L,n_R+1\rangle$ will induce a nonzero order parameter $\varphi_{jR}$.

In the MI regions of $\varphi_{jL}$ (i.e. $\varphi_{jL}=0$), the cluster state is a single Fock state $|n_L,n_R\rangle$.
The quasi-degeneracy of $|n_L,n_R\rangle$ and $|n_L+1,n_R\rangle$ under weak intra-chain hopping, which means that one atom can freely move in the left chain, will result a nonzero order parameter $\varphi_L$.
From the energy quasi-degeneracy relation $E(n_L,n_R)=E(n_L+1,n_R)$, we have the system parameters obeying
\begin{equation}\label{Eq_psi_L}
  \mu=\frac{\Delta}{2}+U n_L, (n_L=0,1,2,...).
\end{equation}
Obviously, the above relation \eqref{Eq_psi_L} between the chemical potential $\mu$ and the bias $\Delta$ well agree with the SF stripes shown Fig.~4 (f).
Similarly, from the energy quasi-degeneracy relation $E(n_L,n_R)=E(n_L,n_R+1)$, we have
\begin{equation}\label{Eq_psi_R}
  \mu=-\frac{\Delta}{2}+U n_R, (n_R=0,1,2,...),
\end{equation}
for the SF stripes of nonzero order parameter $\varphi_{jR}$ shown in Fig.~4 (i).

In addition to the single-atom tunneling along a specific chain, there exists single-atom tunneling between two chains, which corresponds to a nonzero intra-cluster first-order correlation $\textrm{Cor}_{\perp}^{(1)}$.
The inter-chain single-atom tunneling is caused by the quasi-degeneracy of $|n_L,n_R\rangle$ and $|n_L+1,n_R-1\rangle$.
Therefore, from the energy quasi-degeneracy $E(n_L,n_R)=E(n_L+1,n_R-1)$, we have
\begin{equation}\label{Eq_Cor}
  \Delta=U(n_R-n_L),  (n_{L,R}=0,1,2,...)
\end{equation}
As nonzero $\textrm{Cor}_{\perp}^{(1)}$ appears in the region of both $\varphi_{jL} \ne 0$ and $\varphi_{jR} \ne 0$, the corresponding chemical potential $\mu$ is given as,
\begin{equation}\label{Eq_chemical}
\mu = \frac{U}{2}(n_L+n_R),
\end{equation}
with $n_{L,R}=(0,1,2,...)$.
This means that nonzero $\textrm{Cor}_{\perp}^{(1)}$ appears in the vicinity surrounding   $(\Delta^*/U,\mu^*/U)=\left(n_R-n_L,\frac{1}{2}(n_L+n_R)\right)$ with $n_{L,R}=(0,1,2,...)$, see the bright spots in Fig.~4 (b) and (c).
For a given chemical potential $\mu=\frac{1}{2}(n_L+n_R)U$, a sequence of single-atom tunneling and interaction blockade takes place when the bias $\Delta$ increases from negative infinity to positive infinity.
The single-atom tunneling and interaction blockade in the BHL with fixed chemical potential is reminiscent of that of the quantized Bose-Josephson junction with strong interaction~\cite{LeeEPL2008,Cheinet}.
To the best of our knowledge, the single-atom tunneling and interaction blockade in the BHL have never been reported before.

In the second column of Fig.~4, we show the phase diagrams for the case of $\beta=1$.
The parallel SF stripes of $\varphi_{jL}$ or $\varphi_{jR}$ still appear but blur at the quasi-degenerate regions in the vicinity of $(\Delta^*/U,\mu^*/U)$.
Different from the case of large $\beta$, $\varphi_{jL}  \ne 0$ and $\varphi_{jR}  \ne 0$ may coexist in some specific regions.
Correspondingly, due to the increase of $\Jv$, the area of nonzero $\textrm{Cor}_{\perp}^{(1)}$ surrounding $(\Delta^*/U,\mu^*/U)$ extend.

In the first column of Fig.~4, we show the phase diagrams for the case of $\beta=0.1$.
The strong intra-chain tunneling makes the occurrence of inter-chain intra-cluster single-atom tunneling more easy, the regions of $\textrm{Cor}_{\perp}^{(1)} \ne 0$ extend and merge into an entire area.
Correspondingly, due to the strong inter-chain hopping, the properties of $\varphi_{jL}$ and $\varphi_{jR}$ change dramatically.
The parallel SF stripes are tailored and several avoided crossings appear in the vicinity of $(\Delta^*/U,\mu^*/U)$.
The avoided crossings, which have $\textrm{Cor}_{\perp}^{(1)} \ne 0$ and zero order parameters ($\varphi_{jL}=\varphi_{jR}=0$), correspond to the LI phases shown in Fig.~3 (a).
This means that atoms may move freely between the two chains although there is no superfluid along the chain direction.

\section{LANDAU-ZENER DYNAMICS}\label{Sec4}

In this section, we analyze the many-body LZ dynamics in the BHL. In the many-body LZ process, the inter-chain energy bias $\Delta(t)$ is linearly swept from negative to positive or vice versa. The linear sweep of bias is described by $\Delta(t)=\Delta_0+\alpha t$ with $\Delta_0$ being the initial bias and $\alpha$ denoting the sweeping rate. In the first subsection, we show how to apply the Gutzwiller mean-field to the time-evolution problem of our BHL system. In the second subsection, we present the population dynamics in the ground-state sweep and the inverse sweep, respectively. In the ground-state sweep, the initial state is the ground state, while in the inverse sweep, the initial state is the highest excited state.

\begin{figure}[!htp]
  \includegraphics[width=\columnwidth]{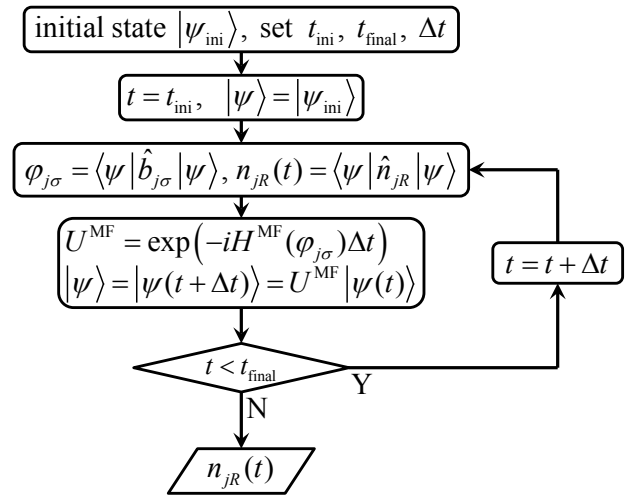}
  \caption{\label{Fig_TEP}The numerical simulation procedure for the time-evolution of Bose-Hubbard ladder system via our cluster Gutzwiller mean-field method.}
\end{figure}

\subsection{Time-evolution problem}

The time-evolution obeys the Schr\"{o}dinger equation
\begin{equation}\label{Eq_Sch}
  i\hbar \frac{d}{dt}\ket{\Psi(t)}=\hat{H}(t)\ket{\Psi(t)},
\end{equation}
where $\hat{H}(t)$ is the original time-dependent Hamiltonian~\eqref{Eq_Hamiltonian}.
By applying the dynamical Gutzwiller mean-field method, the time-evolution is described by the dynamical Gutzwiller equations
\begin{equation}\label{Eq_dynamics_MF}
i\hbar \frac{d}{dt}\ket{\Psi^\mathrm{GA}(t)}=\hat{H}^\mathrm{MF}(t)\ket{\Psi^\mathrm{GA}(t)},
\end{equation}
where $\hat{H}^\mathrm{MF}(t)$ is the time-dependent mean-field Hamiltonian and $\ket{\Psi^\mathrm{GA}(t)}=\prod_{j}\ket{\Psi(t)}_j$ denotes the time-dependent Gutzwiller ansatz.
Here, the single-cluster state reads as
\begin{equation}\label{Eq_wavefun_j_t}
\ket{\Psi(t)}_j=\sum_{N=0}^{N_\mathrm{max}}\sum_{m=-N}^{N}f_{N,m}^{(j)}(t)\ket{N,m}_j.
\end{equation}

Similar to determining the ground states, the time-dependent mean-field Hamiltonian $\hat{H}^\mathrm{MF}(t)$ can be decoupled as a sum of single-cluster Hamiltonians. Therefore, the dynamical Gutzwiller equations~\eqref{Eq_dynamics_MF} can be simplified to the single-cluster equations
\begin{equation}\label{Eq_dynamics j}
  i\hbar \frac{d}{dt}\ket{\Psi_j(t)}=\hat{H}_j^\mathrm{MF}(t)\ket{\Psi_j(t)},
\end{equation}
with the time-dependent single-cluster Hamiltonian
\begin{eqnarray}\label{Eq_MFHam_j}
  \hat{H}_j^\mathrm{MF}(t)&=&-\Jp\sum_{\sigma,k=j\pm1}\left(
  \varphi_{k\sigma}\hatd{b}_{j\sigma}+\varphi_{k\sigma}^*\hat{b}_{j\sigma}
  -\mathrm{Re}\left[\varphi_{j\sigma}^*\varphi_{k\sigma}\right]\right)\nonumber\\
  &&-\Jv\left(\hatd{b}_{jL}\hat{b}_{jR}+\mathrm{h.c.}\right)
  +\frac{U}{2}\sum_{\sigma}\hat{n}_{j\sigma}\left(\hat{n}_{j\sigma}-\hat{1}\right)\nonumber\\
  &&-\frac{\Delta(t)}{2}(\hat{n}_{jR}-\hat{n}_{jL})-\mu\sum_{\sigma}\hat{n}_{j\sigma},
\end{eqnarray}
in which the time-dependent order parameters are given as $\varphi_{j\sigma}(t)=\langle \Psi_j(t)|\hat b_{j\sigma}|\Psi_j(t) \rangle$.
Substituting Eq.~\eqref{Eq_wavefun_j_t} and Eq.~\eqref{Eq_MFHam_j} into Eq.~\eqref{Eq_dynamics j}, one can obtain the following differential equations for the expansion coefficients
\begin{widetext}
\begin{eqnarray}\label{Eq_dynamics}
  i\hbar \frac{d}{dt}f_{N,m}(t)=
  &-&\frac{\Jp}{\sqrt{2}}\phi_{jL}(t)\sqrt{N+m}f_{N-1,m-1}(t) -\frac{\Jp}{\sqrt{2}}\phi_{jR}(t)\sqrt{N-m}f_{N-1,m+1}(t)\nonumber\\ &-&\frac{\Jp}{\sqrt{2}}\phi_{jL}(t)^*\sqrt{N+m+2}f_{N+1,m+1}(t)
  -\frac{\Jp}{\sqrt{2}}\phi_{jR}(t)^*\sqrt{N-m+2}f_{N+1,m-1}(t)\nonumber\\ &-&\frac{\Jv}{2}\sqrt{N+m}\sqrt{N-m+2}f_{N,m-2}(t)
  -\frac{\Jv}{2}\sqrt{N+m+2}\sqrt{N-m}f_{N,m+2}(t)\nonumber\\ &+&\left[\frac{U}{4}\left(N^2+m^2-2N\right)+\frac{\Delta(t)}{2}{m}-\mu{N}
  +\Jp\mathrm{Re}\left[\varphi_{jL}(t)^*\phi_{jL}(t)+\varphi_{jR}(t)^*\phi_{jR}(t)\right]\right]f_{N,m}(t),
\end{eqnarray}
\end{widetext}
with the time-dependent order parameters
\begin{eqnarray}
  \phi_{j\sigma}(t)&=& \varphi_{j+1,\sigma}(t)+\varphi_{j-1,\sigma}(t), \\
  \varphi_{jL}(t)&=&\sum_{N,m}\sqrt{\frac{N+m+2}{2}}f_{N,m}^{(j)*}(t)f_{N+1,m+1}^{(j)}(t), \\
  \varphi_{jR}(t)&=&\sum_{N,m}\sqrt{\frac{N-m+2}{2}}f_{N,m}^{(j)*}(t)f_{N+1,m-1}^{(j)}(t).
\end{eqnarray}
By using the fourth-order Ronge-Kutta method, we simulate the dynamics obeying Eq.~(\ref{Eq_dynamics}). The flow chart for the numerical procedure is shown in Fig.~\ref{Fig_TEP}. Given the parameters $\Jp$, $\Jv$, $U$, the initial bias $\Delta_0$, the sweeping rate $\alpha$, and the initial state, the time-dependent order parameters should be estimated by the instantaneous states step by step. That is, for a specific time step, based upon the current state and the current order parameters, we need estimate not only the time-dependent state but also the time-dependent order parameters for the next time step.

\begin{figure*}[!htp]
  \includegraphics[width=17.2 cm]{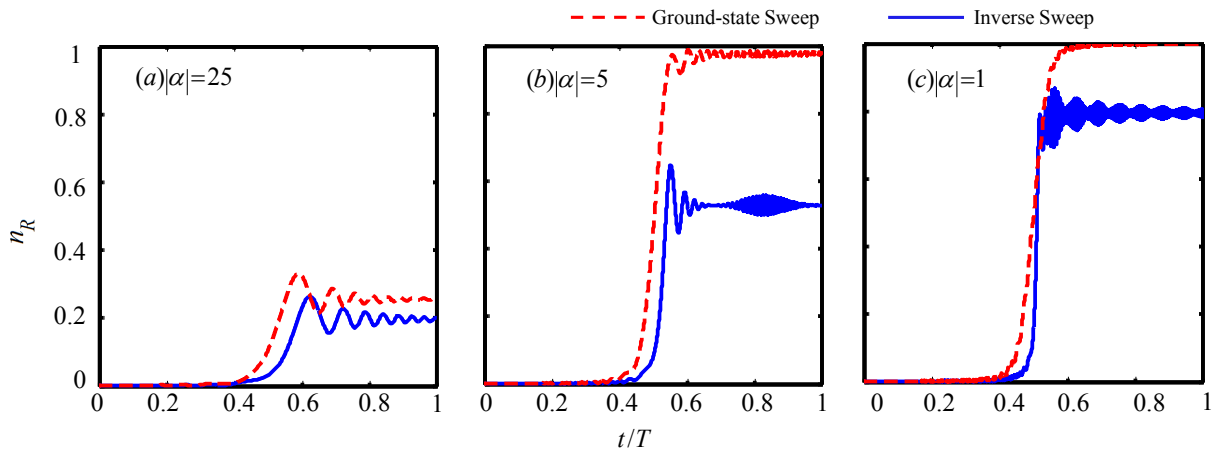}
  \caption{\label{Fig_dynamics} (Color online) The many-body Landau-Zener dynamics in the Bose-Hubbard ladder for different sweeping rates $|\alpha|$. Here, $n_R(t)$ stands for the transfer fraction, the cutoff of the maximum particle number is fixed as $N_{\mathrm{max}}$=6. The other parameters are chosen as: $|\Delta_0|=50$, $\mu=3$, $U=0.5$, $\Jp=0.25$, $\Jv=1$.
  Red-dashed lines and blue-solid lines correspond to the ground-state sweep and the inverse sweep, respectively.
  (a) For a large sweep rate, $|\alpha|=25$, the transfer fractions $n_R(T)$ for both the ground-state sweep and the inverse sweep are much smaller than $1$. The difference between the the two sweeps is small.
  (b) For an intermediate sweep rate, $|\alpha|=5$, the transfer fractions $n_R(T)$ for the ground-state sweep increases to almost $1$, and $n_R(T)$ for the inverse sweep is still far below $1$.
  (c) For a small sweep rate, $|\alpha|=1$, the transfer fractions $n_R(T)$ for the ground-state sweep reaches $1$, while $n_R(T)$ for the inverse sweep is still below $1$. This means that the ground-state sweep evolves adiabatically, while the inverse sweep evolves non-adiabatically.}
\end{figure*}

\subsection{Population dynamics}

We consider two typical sweep processes: the ground state sweep and the inverse sweep.
In the ground-state sweep, the system is prepared in the ground state of all particles in the lower chain,
and the initial bias between left and right chains is set as $\Delta_0=-50$ and then the bias $\Delta(t)$ is linearly swept from $\Delta_0$ to $-\Delta_0$ with the sweep rate $\alpha=-2\Delta_0/T >0$.
In the inverse sweep, the system is prepared in the highest excited state of all particles in the higher chain,
and the initial bias between left and right chains is set as $\Delta_0=50$ and then the bias $\Delta(t)$ is linearly swept from $\Delta_0$ to $-\Delta_0$ with the sweep rate $\alpha=-2\Delta_0/T <0$.
Here $T$ is the total sweep time.
For convenience, we assume that the initial state for both two sweep processes is the state of all atoms in the left chain.
To show the many-body LZ dynamics, we calculate the transfer fraction $n_R(t)=N_R(t)/N$, which is the fraction of the particles in the right chain at a given time $t$.
Obviously, the bias $\Delta(t)$ vanishes at time $t=T/2$, which corresponds to an instantaneous symmetric BHL.

If there are no intra-chain hopping and no on-site interaction, i.e. $\Jp=0$ and $U=0$, the physical picture for the many-body LZ dynamics is as same as the one for the conventional two-level LZ problem.
This means, the final transfer efficiency is given by the conventional LZ formula $n_R(+\infty)=1-\exp(-2\pi\Jp^2 / \hbar|\alpha|)$ and there is no significant difference between the ground-state and inverse sweeps.
However, taking into account the on-site interaction and the intra-chain hopping, the many-body LZ dynamics becomes very different from the conventional two-level LZ problem.
Below, we analyze the many-body LZ dynamics for the on-site interaction $U=0.5$, the inter-chain hopping $\Jv=1$, the intra-chain hopping $\Jp=0.25$, and different sweep rates $\alpha$.

Independent on the sweep rate $\alpha$, significant population transfers from the left chain to the right chain appear around the time $t=T/2$.
This significant population transfer between the two chains is caused by the avoided level crossing in the vicinity of the bias $\Delta(t)= 0$.
However, the transfer fraction $n_R(t)=N_R(t)/N$ sensitively depends on the sweep rates, the physical parameters and the initial states.
In particular, for slow sweep rates, there appear significant difference of the final transfer fraction for the ground-state sweep and the inverse sweep, see Fig.~\ref{Fig_dynamics}.

For a large sweep rate, $|\alpha|=25$, both the ground-state sweep ($\alpha=+25$) and the inverse sweep ($\alpha=-25$) are non-adiabatic, see Fig.~\ref{Fig_dynamics} (a).
The dynamics of the ground-state sweep and the inverse sweep is very similar.
The transfer fraction $n_R(t)$ rapidly increase around $\Delta(t)=0$ and then keep oscillates around a specific value.
The final transfer fraction $n_R(T)$ is much below 1 because of the non-adiabatic evolution under fast sweeps.

For an intermediate sweep rate, $|\alpha|=5$, the non-adiabatic excitation in the ground-state sweep is not very significant, while the non-adiabatic excitation in the inverse sweep is very significant, see Fig.~\ref{Fig_dynamics} (b).
After the system goes through the avoided level crossing region around $\Delta(t)=0$, the transfer fraction for the ground-state sweep ($\alpha=+5$) is very close to 1 and its oscillation amplitude is very small.
While in the inverse sweep ($\alpha=-5$), the final transfer fraction is much below 1 and the corresponding oscillation amplitude is much larger than the one for the ground-state sweep.

For a small sweep rate, $|\alpha|=1$, the ground-state sweep undergoes adiabatic evolution and but the inverse sweep still show significant non-adiabatic excitations, see Fig.~\ref{Fig_dynamics} (c).
In the ground-state sweep ($\alpha=+1$), there is no significant oscillations in the transfer fraction and the final transfer fraction is almost the perfect limit $n_R(T)=1$, which means that all particles in the left chain can be completely transferred into the right chain.
However, in the inverse sweep ($\alpha=-1$), the final transfer fraction is still much below 1 and the oscillation amplitude is still very significant, which indicates that there still exist significant non-adiabatic excitations.

The adiabaticity breakdown in the inverse sweep qualitatively explains the recent experimental observation~\cite{Chen2011, Kasztelan2011}.
The observed adiabaticity breakdown, which can not be found in the conventional two-level LZ problem, is a result of the inter-particle interaction.
Due to the inter-particle interaction, swallow-tail-shaped loop structures~\cite{WuPRA2000, LiuPRA2002}, which correspond to the macroscopic quantum self-trapping in mean-field models~\cite{SmerziPRL1997, RaghavanPRA1999, KuangPRA2000, LeePRA2001, LeePRA2004, AlbiezPRL2005, GatiJPB2007, LeePRL2009, LeeFOP2012}, may appear in the energy spectrum for our BHL system.
Unlike the conventional two-level LZ problem, whose energy-level structures for the ground state and the highest-excited state are similar, the energy-level structures for the ground state and the highest-excited state of our BHL system are very different.
Because of their different energy-level structures, the ground-state sweep and the inverse sweep show different adiabatic/non-adiabatic dynamics.

\section{CONCLUSION AND DISCUSSION}\label{Sec5}

In summary, we present a cluster Gutzwiller mean-field approach to explore the static and dynamical behavior of the BHL, which can be experimentally realized by loading Bose atoms into a double-well optical superlattice potential.
In our mean-field treatment, the wavefunction of the whole system is assumed in form of the Gutzwiller ansatz, the two sites in each double-well unit are packed as a cluster and the inter-cluster hopping is decoupled by using the conventional mean-field approximation.
Through implementing the numerical self-consistent procedure, for both unbiased and biased BHLs, we obtain the ground states and give the phase diagram by calculating the order parameters.

For an unbiased BHL, if the intra-chain hopping is much stronger than the inter-chain hopping (i.e. $\beta<<1$), there appear several exotic loophole-shaped insulator regions of the half-integer filling numbers, which lie between the conventional MI lobes of integer filling numbers.
As $\beta$ increases, the loophole-shaped insulator regions gradually shrink and disappear.
Differently, if the inter-chain hopping is much stronger than the intra-chain hopping (i.e. $\beta>>1$), the unbiased BHL system can be regarded as two single BH chains and the corresponding phase diagram is almost as same as the one for a single BH chain.

For a biased BHL, single-atom tunneling and interaction blockade appear if the hopping terms are weak enough to be treated as perturbations.
We present an analytical interpretation for the single-atom tunneling and interaction blockade based upon the quasi-degeneracy of different Fock states for the considered cluster.
If the inter-chain hopping is much stronger than the intra-chain one, there appear exotic LI phases with no superfluids along the chain direction but nonzero inter-chain coherence.

In further, we analyze the many-body LZ process of the BHL, in which the inter-chain bias is linearly swept from positive to negative or vice versa.
We consider two different sweeps: the ground-state sweep and the inverse sweep.
In the ground-state sweep, the initial state is the ground state and the final transfer fraction can reach 1 if the sweep rate is small enough.
While in the inverse sweep, whose initial state is the highest excited state, there still exist significant non-adiabatic excitations when the corresponding ground-state sweep obeys adiabatic evolution.
The breakdown of adiabaticity in the inverse sweep, which are well consistent with the recent experimental observations~\cite{Chen2011, Kasztelan2011}, is a result of the swallow-tail-shaped loop structures induced by inter-particle interaction~\cite{WuPRA2000, LiuPRA2002}.

In recent, for ultracold atoms in optical lattices, artificial gauge fields have been realized by lattice shaking technique~\cite{Struck2013} or laser-induced tunneling~\cite{Atala2014}.
The artificial gauge fields, which allow one to generate spin-orbit couplings and effective magnetic fields, opens a new path to explore quantum Hall effect and topological phases of matters.
Our cluster Gutzwiller mean-field approach can also be extended to investigate the bosonic ladders in the presence of an artificial magnetic field~\cite{Aidelsburger2011, Atala2014, Aidelsburger2014, Greschner2015, Tokuno2014, Keles2015, XPLi2015, Hugel2014}, such as the observation of chiral currents~\cite{Atala2014}, the measurement of Chern number in Hofstadter bands~\cite{Aidelsburger2014, Hugel2014}, and the two-leg Bose-Hubbard ladder under a magnetic flux~\cite{Tokuno2014,Keles2015}.
In addition, our cluster Gutzwiller mean-field approach may also use to explore the non-equilibrium dynamics of two coupled one-dimensional Luttinger liquids~\cite{Foini2015} and the dynamical instability of interacting bosons in disordered lattices~\cite{Buonsante2015}.

\section*{Acknowledgements}

H. Deng, H. Dai and J. Huang made equal contributions for this work.
This work is supported by the National Basic Research Program of China (NBRPC) under Grant No. 2012CB821305, the National Natural Science Foundation of China (NNSFC) under Grants No. 11374375 and No. 11465008, and the PhD Programs Foundation of Ministry of Education of China under Grant No. 20120171110022.


\end{document}